\documentclass{amsart}
\usepackage{amsmath}
\usepackage{amsfonts}
\usepackage{graphics}
\usepackage{amssymb}
\usepackage{amscd}
\usepackage[all]{xy}
\usepackage{latexsym}
\usepackage{graphicx}
\usepackage{multirow}
\usepackage{geometry}

\theoremstyle{plain}
\newtheorem{lem}{\bf Lemma}[section]

\newtheorem{thm}{\bf Theorem}[section]
\newtheorem{cor}{\bf Corollary}[section]
\newtheorem{ex}{\bf Example}[section]

\numberwithin{equation}{section} \numberwithin{figure}{section}

\renewcommand*{\to}{\rightarrow}
\renewcommand*{\bar}[1]{\overline{#1}}

\newcommand{\Td}{\operatorname{Td}}

\newcommand{\ch}{\operatorname{ch}}

\newcommand{\ind}{\operatorname{ind}}
\newcommand{\Gr}{\operatorname{Gr}}

\newcommand{\coker}{\operatorname{coker}}

\newcommand{\mb}[1]{\mathbb{#1}} % Field

\newcommand{\hs}{\mathcal{H}}% Hilbert space
\newcommand{\mc}[1]{\mathcal{#1}}

\begin{document}
\large \setcounter{section}{0}

\title{ Moduli spaces and Grassmannian}\thanks {The work was partially supported by NSF grant DMS-0805989}
\author{Jia-Ming (Frank) Liou}
\address{Max Planck Institut f\"{u}r Mathematik\\
Vivatsgasse 7\\
Bonn, 53111, Germany\\fjmliou@gmail.com }

\author{A. Schwarz}
\address{Department of Mathematics\\
University of California\\
Davis, CA 95616, USA\\ schwarz@math.ucdavis.edu}

\maketitle

\begin{abstract}\large
We calculate the homomorphism of the cohomology induced by the
Krichever map of moduli spaces of curves into infinite-dimensional
Grassmannian. This calculation can be used to compute the homology
classes of cycles on moduli spaces of curves that are defined in
terms of Weierstrass points.
\end{abstract}



\allowdisplaybreaks

%\tableofcontents

\section{Introduction}

We study the relation between the topology of Sato Grassmannian and
the topology of the moduli space of compact complex curves. The Sato
Grassmannian (or, better to say the Segal-Wilson \cite{GS} version
of Sato Grassmannian) associated with a polarized Hilbert space
$\hs=\hs_{+}\oplus\hs_{-}$ is an infinite dimensional Banach
manifold $\Gr(\hs)$  modeled on the space of compact operators from
$\hs_{-}$ to $\hs_{+}$. Its path components are parametrized by the
set of integers.  The cohomology  of each path component can be
identified with the cohomology of the infinite classical
Grassmannian \cite{PS}. Let $\Gr_{d}(\hs)$ be a path connected
component of $\Gr(\hs)$, where $d$ is an integer. The cohomology
ring of $H^{*}(\Gr_{d}(\hs))$ is isomorphic to the polynomial ring
$\mb C[c_{1},c_{2},\dots]$ with variables $c_{k}$ whose degrees are
$2k$.  Since $S^{1}$ acts naturally on $\Gr(\hs)$, we can also
consider the $S^{1}$-equivariant cohomology of $\Gr(\hs)$.

The moduli space $\widehat{\mc F}_{g,h}$ is the space of quintuples
$(C,p,z,L,\phi)$ where $C$ is a compact complex curve, $z$ is a
local coordinate in the disk $D$ centered at the point $p\in C$, and
$L$ stands for a line bundle over $C$ having a trivialization $\phi$
over $D$. This space can be mapped into the Sato Grassmannian
$\Gr(\hs)$ by means of Krichever construction sending
$x=(C,p,z,L,\phi)\in\widehat{\mc F}_{g,h}$ to the closed subspace of
$\hs$ consisting of functions $f:S^{1}\to\mb C$ that can be obtained
as restrictions  of holomorphic sections of $L$ over $C\backslash
D$. (See, for example \cite {MM}.) This construction determines an
embedding $k:\widehat{\mc F}_{g,h}\to\Gr(\hs)$. The image of
$\widehat{\mc F}_{g,h}$ in $\Gr(\hs)$ via $k$ is called the
Krichever locus. The continuous map $k:\mc F_{g,h}\to \Gr(\hs)$
induces a homomorphism on cohomology ring $k^{*}:H^{*}(\Gr(\hs))\to
H^{*}(\widehat{\mc F}_{g,h})$. The group $U(1)=S^1$ acts in natural
way on $\widehat{\mc F}_{g,h}$ and on $\Gr(\hs)$; this action
commutes with Krichever map hence we can talk about corresponding
homomorphism of equivariant cohomology. We analyze the induced map
$k^{*}$ both for conventional cohomology and equivariant cohomology.
We express this map in terms of lambda-classes, introduced by
Mumford \cite {Mum}, and their generalizations.

The cohomology of Sato Grassmannian can be represented by
finite-codimension subvarieties that are called Schubert cycles.
{\footnote {The statement  that  a submanifold of oriented manifold
specifies a cohomology class of dimension equal to the codimension
of submanifold is well known in finite-dimensional case, but its
precise formulation and proof in infinite-dimensional case are
non-trivial. However, in the situations we consider it is possible
to justify our considerations representing infinite-dimensional
manifolds as limits of finite-dimensional ones. }}The intersections
of Schubert cycles with Krichever locus can be described as cycles
on moduli spaces that are  defined in terms of Weierstrass points
\cite {AK}. Our calculations can be interpreted as calculations of
(co)homology classes of these cycles provided that  the Schubert
cycle and Krichever locus are in general position.

The results of the present paper should be important in the analysis
of Grassmannian string theory suggested in \cite {AS} as a version
of nonperturbative string theory. The ideas of \cite {AS} should be
combined with ideas of \cite {SZ}; this leads to the analysis of
BV-algebra  of equivarianted chains on Grassmannian. The present
paper  is a  first step in this direction.

\section{Preliminaries}

\subsection{Sato Grassmannian}
A semi-infinite structure on an infinite dimensional separable
Hilbert space $\hs$ is a triple $(\hs_{+},\hs_{-},\kappa)$, where
$\hs_{+}$ and $\hs_{-}$ are infinite dimensional closed subspace of
$\hs$ with $\hs=\hs_{+}\oplus\hs_{-}$ and $\kappa:\hs\to\hs$ is an
invertible map so that $\kappa:\hs_{\pm}\to\hs_{\mp}$. A polarized
Hilbert space is a Hilbert space $\hs$ together with a semi-infinite
structure $(\hs_{+},\hs_{-},\kappa)$ on it. A polarized Hilbert
space is denoted by $(\hs,\kappa)$. Given a polarized Hilbert space
$(\hs,\kappa)$, denote the orthogonal projections from $\hs$ onto
$\hs_{\pm}$ by $\pi_{\pm}$ respectively.
%A morphism
%$T:(\hs,\kappa)\to(\hs',\kappa')$ between two polarized Hilbert
%spaces is a linear isometry $T$ from $\hs$ to $\hs'$ so that
%$T(\hs_{\pm})\subset \hs_{\pm}'$ and $\kappa'\circ T=T\circ\kappa$.
%The class of polarized Hilbert spaces form a category.

The Sato Grassmannian\footnote{We use the version of Sato
Grassmannian defined by Segal and Wilson.} $\Gr(\hs)$ associated
with a polarized Hilbert space $(\hs,\kappa)$ is the set of all
closed subspaces $W$ of $\hs$ such that the orthogonal projection
$\pi_{-}|_{W}:W\to\hs_{-}$ is a Fredholm operator and the orthogonal
projection $\pi_{+}|_{W}:W\to\hs_{+}$ is a compact operator
{\footnote {Sometimes it is convenient to use Hilbert-Schmidt
operators instead of compact operators in the definition of
Grassmannian.  Our calculations can be applied to this modification
of Grassmannian.}}.

For each $W\in\Gr(\hs)$, let $U_{W}$ be the set of all closed
subspaces which are graphs of compact operators from $W$ into
$W^{\perp}$. In other words, $V\in U_{W}$ if and only if $V$
consists of points of the form $w+Kw, w\in W$ for some compact
operator $K:W\to W^{\perp}$. Define $\varphi_{W}(V)=K$, where
$V=W+KW$ in $U_{W}$. Then $\varphi_{W}:U_{W}\to \mc K(W,W^{\perp})$
is a bijection for each $W\in\Gr(\hs)$.  (Here $\mc K(V,V')$ stands
for the space of compact operators from a Banach space $V$ to a
Banach space $V'$). The family $\{(U_{W},\varphi_{W})\}$ gives
$\Gr(\hs)$ a Banach manifold structure modelled on $\mc
K(\hs_{+},\hs_{-})$. A typical example of a polarized Hilbert space
is $L^{2}(S^{1})$ together with the standard semi-infinite structure
defined as follows. The subspaces $L^{2}(S^{1})_{+}$ and
$L^{2}(S^{1})_{-}$ of $L^{2}(S^{1})$ are the closed subspaces of
$L^{2}(S^{1})$ spanned by $\{z^{i}:i\geq 0\}$ and $\{z^{j}:j<0\}$
respectively and form an orthogonal direct sum of $L^{2}(S^{1})$.
The standard semi-infinite structure on $L^{2}(S^{1})$ is the map
$\kappa(f)(z)=\frac{1}{z}f\left(\frac{1}{z}\right)$. One can see
that $\kappa$ maps $L^{2}(S^{1})_{\pm}$ into $L^{2}(S^{1})_{\mp}$.
The standard Sato Grassmannian $\Gr(L^{2}(S^{1}))$ is the Sato
Grassmannian associated with the standard polarized Hilbert space
$(L^{2}(S^{1}),\kappa)$. From now on, we will assume
$\hs=L^{2}(S^{1})$. A point $W$ in $\Gr(\hs)$ is said to have
virtual index $d$ if the Fredholm operator $\pi_{-}|_{W}$ has index
$d$. The set of all points of $\Gr(\hs)$ consisting of virtual index
$d$ forms a submanifold of $\Gr(\hs)$; it is denoted by
$\Gr_{d}(\hs)$. The manifolds  $\{\Gr_{d}(\hs)\}_{d\in\mb Z}$ are
connected components of $\Gr(\hs)$.

\subsection{Moduli Spaces and the Krichever Map}

We denote by $\mc M_g$ the moduli space of complex curves of genus
$g$ (of Riemann surfaces of genus $g$). As a set this is a set of
all equivalence classes of compact smooth complex curves of genus
$g$.  A rigorous definition of moduli space $\mc M_{g}$ is
complicated, because complex curves can have non-trivial
automorphisms. This means that the moduli space should be regarded
as an orbifold or as a stack. To avoid these complications we can
work instead with families of curves . {\footnote {Recall that a
family of curves of genus $g$ with base $B$ is a holomorphic map
$p:E\to B$ that can be considered as locally trivial fibration with
curves of genus $g$ as fibers. }} A complex curve of genus $g$ with
$n$ marked points is a collection $(C,p_{1},\dots,p_{n})$, where $C$
is a compact complex curve of genus $g$ and $(p_{1},\dots,p_{n})$ is
an $n$-tuple of distinct points on $C$. A morphism from
$(C,p_{1},\dots,p_{n})$ to $(C',p_{1}',\dots,p_{n}')$ is a
holomorphic map $\varphi:C\to C'$ such that $\varphi(p_{j})=p_{j}'$
for $1\leq j\leq n$. The moduli space of  complex curves with $n$
marked points $\mc M_{g,n}$ consists of all isomorphism classes of
compact complex curves of genus $g$ with $n$ marked points. It is
obvious that $\mc M_{g,0}=\mc M_{g}$. Again we can work with
families instead of moduli spaces. Similarly we can define other
moduli spaces.

The moduli space $\widehat{\mc M}_{g}$ is defined as a space of triples $(C,p,z)$, where $C$ is a compact
complex curve of genus $g$ with a point $p$ and a map $z:D\to \mb D$
is an isomorphism from a closed set $D$ into the closed unit disk
$\mb D=\{z\in\mb C:|z|\leq 1\}$ obeying $z(p)=0$ . The moduli space of quintuples
$(C,p,z,L,\phi)$, where $(C,p,z)$ specifies a point of $\widehat{\mc
M}_{g}$ and $L$ is a line bundle over $C$ together with a local
trivialization $\phi$ over $D$ will be  denoted by $\widehat{\mc
F}_{g,h}$. We also denote by $\mc F_{g,h}$ the moduli space of triples
$(C,p,L)$, where $(C,p)\in \mc M_{g,1}$ and $L$ is a line bundle
over $C$ of degree $h$.The moduli space of pairs
$(C,L)$, where $C\in\mc M_{g}$ and $L$ is a line bundle over $C$ of
degree $h$ will be denoted by   $\mc P_{g,h}$ .

The moduli spaces $\widehat{\mc M}_{g}$ and
$\widehat{\mc F}_{g,h}$ can be embedded into the standard Sato
Grassmannian $\Gr(\hs)$.

Let $(C,p,z,L,\varphi)$ be a point in $\widehat{\mc F}_{g,h}$.
Identify $D$ with the closed unit disk $\mb D$ and its boundary
$\partial D$ with $S^{1}$ via $z$. Let $H^{0}(C\setminus D,L)$ be
the space of holomorphic sections of $L$ over $C\setminus D$. Let
$k(x)$ be the closed subspace of $L^{2}(S^{1})$ consisting of
functions $f$ with the property that there exists $s\in
H^{0}(C\setminus D,L)$ such that $f=s|_{S^{1}}$. One can show that
$k(x)\in\Gr(\hs)$ (\cite{GS}). Moreover, $\ker\pi_{-}|_{k(x)}$ and
$\coker\pi_{-}|_{k(x)}$ can be identified with
$\ker\bar{\partial}_{L}=H^{0}(C,L)$ and
$\coker\bar{\partial}_{L}=H^{1}(C,L)$ respectively. By the
Riemann-Roch theorem,
\begin{equation*}
\ind k(x)=h^{0}(C,L)-h^{1}(C,L)=h-g+1,
\end{equation*}
where $h^{i}(C,L)=\dim H^{i}(C,L)$ for any line bundle $L$. The map
\begin{equation}\label{Kh}
k:\widehat{\mc F}_{g,h}\to \Gr_{d}(\hs)
\end{equation}
where $d=h-g+1$ is called the Krichever map . It is a continuous
embedding. Similarly, one can construct  continuous embeddings
\begin{equation}\label{Kq}
k_{q}:\widehat{\mc M}_{g}\to \Gr_{d_{q}}(\hs)
\end{equation}
by defining $k_{q}(C,p,z)=k(C,p,z,K_{C}^{\otimes q},dz^{\otimes
q})$, where $K_{C}$ is the canonical line bundle over $C$,
$h_{q}=q(2g-2)$ for $q\geq 1$, and $d_{q}=h_{q}-g+1$. (One can say
that these embeddings are obtained as compositions of Krichever map
and natural embeddings of $\widehat{\mc M}_{g}$ into $\widehat{\mc
F}_{g,h}$.)

\subsection{The Equivariant Cohomology}
Let $G$ be a topological group and $X$ be a $G$-space. The
equivariant cohomology of $X$ is defined to be
\begin{equation*}
H_{G}^{*}(X)=H^{*}(EG\times_{G}X),
\end{equation*}
where $EG$ is a contractible $G$-space such that $G$ acts freely on
$EG$ and $EG\times_{G}X$ denotes the quotient space of $EG\times X$
modulo the relation $(h\cdot g,x)\sim (h,g\cdot x)$. ( This
definition works for any group of coefficients, but we always
consider the cohomology with coefficients in $\mathbb{C}$.) When $X$
is a point, $H_{G}^{*}(pt)$ is the cohomology $H^{*}(BG)$ of the
classiying space $BG=EG/G$ . When $G$ acts freely on $X$,
$H_{G}^{*}(X)$ is simply $H^{*}(X/G)$. The equivariant cohomology
$H_{G}^{*}(X)$ is an algebra over $H_{G}^{*}(pt)$.

If $G$-space $X$ is an orientable manifold then for every $G$-
invariant cycle $Z$ of codimension $r$ in $X$  one can construct an
$r$-dimensional   equivariant  cohomology class $[Z]$. We will say
that this class is dual to $Z$ (the construction generalizes
Poincare duality). If $Y$ is a $G$-invariant submanifold of $X$ and
$G$-invariant cycle $Z$ in $X$ is in general position with respect
to $Y$ (the codimension of $Z\cap Y$ in $Y$ is equal to intersection
of $Z$ in $X$) then
\begin{equation}
\label{int} i^*[Z]=[Z\cap Y]
\end{equation}
where $i^*$ denotes the  homomorphism  $H_G^*(X)\to H_G^*(Y)$
induced by the embedding $i:Y\to X$.

\begin{ex}
Let $S^{\infty}$ be an infinite-dimensional sphere (understood as
the direct limit of finite-dimensional spheres with respect to maps
$S^{2n-1}\to S^{2n+1}$ induced by natural embeddings $\mathbb{C}^n
\subset \mathbb{C}^{n+1}$ or as the unit sphere in an infinite
dimensional complex Hilbert space $\hs$). It is a contractible space
with a free $S^{1}$ action. Hence the classifying space $BS^{1}$ of
$S^{1}$ is the infinite dimensional projective space $\mb
P^{\infty}$.  The equivariant cohomology ring
$H_{S^{1}}(pt)=H^{*}(\mb P^{\infty})$ of a point is the polynomial
ring $\mb C[u]$, where $u$ is a degree $2$ element  in $H^{*}(\mb
P^{\infty})$. Using this this statement  one obtains that the
equivariant cohomology ring $H_\mathbb{T}(pt)$ where
$\mathbb{T}=(S^1)^n$ is an $n$-dimensional torus is a polynomial
ring $\mb C [u_1,\dots,u_n]$.
\end{ex}

%\begin{thm}
%Let $d$ be an integer. The group $S^{1}$ acts on $\hs$ and induces
%an action on $\Gr(\hs)$ so that the equivariant cohomology of
%$\Gr_{d}(\hs)$ is a free polynomial algebra generated by $\{c_{r}\}$
%over $\mb R[u]$, where $\deg c_{j}=2j$.
%\end{thm}
One says that a $G$-space $X$ is equivariantly formal if its
equivariant cohomology is a free module over $H_{G}^{*}(pt)$. There
exist numerous conditions that guarantee equivariant formality (see
\cite {GKM}); for us it is sufficient to know that  among these conditions is  vanishing of odd-dimensional cohomology.

Let us suppose that $G=\mb T$ is a torus and the action of $\mb T$
on $X$ is equivariantly formal. Then the restriction map $H_{\mb
T}(X)\to H_{\mb T}(F)$  where  $F$ is the set of fixed points of
torus action is injective. Hence to  calculate the cohomology ring
$H_{\mb T}(X)$ one should describe the image of this map. This can
be done \cite {GKM}. We will formulate the answer in the case when
$X$ is a non-singular algebraic variety, $F$ is finite, the action
of $\mb T$ can extended to an algebraic action of algebraic torus
$\mb T^{al}$ and $\mb T^{al}$  has only a finite number of orbits of
complex dimension $1$.  We define $X_1$ as a union of these orbits
and $F$ (a union of orbits of dimension $\leq 1$). Then the image of
restriction map can be characterized as the kernel of the
homomorphism {\footnote {This homomorphism is defined as the
boundary homomorphism in the exact sequence of the pair $(X_1,F)$.}}
$H_{\mb T}(F)\to H_{\mb T} (X_1,F))$ (GKM-theorem, \cite {GKM}).
Notice, that in the conditions of GKM theorem we can calculate not
only equivariant cohomology with respect to the torus $\mb T$, but
also equivariant cohomology with respect to any subtorus $\mb
{T}'\subset \mb {T}$.
%map defined by $\mc K_{\pi}^{\otimes q}$, where $d_{q}=(2q-1)(g-1)$.
%Now, we want to compute the induced cohomology map on the
%equivariant cohomology:
%\begin{equation*}
%H_{S^{1}}^{*}(\widehat{\mc M}_{g}).
%\end{equation*}

The moduli space $\widehat{\mc M}_{g}$ has a natural free $S^{1}$-
action $S^{1}\times \widehat{\mc M}_{g}\to \widehat{\mc M}_{g}$
defined by $(\lambda,(C,p,z))\mapsto (C,p,\lambda z)$. Let us
consider the forgetful map
\begin{equation*}
F:\widehat{\mc M}_{g}\to\mc M_{g,1}
\end{equation*}
defined by $F(C,p,z)=(C,p)$. The moduli space $\widehat{\mc M}_{g}$
is homotopy equivalent to the moduli space $\mc M_{g,1}'$, where
$\mc M_{g,1}'$ is the moduli space of triples $(C,p,v)$, where
$(C,p)\in \mc M_{g,1}$ and $v$ is a nonzero tangent vector to $C$ at
$p$. Hence $F$ is homotopy equivalent to the map:
\begin{equation*}
F':\mc M_{g,1}'\to\mc M_{g,1},
\end{equation*}
where $F'(C,p,v)=(C,p)$ is the forgetful map. Since the fiber of
$F'$ is homotopy equivalent to $S^{1}$, we have the following
identification:
\begin{equation}\label{ID'}
\bar{F}':\mc M_{g,1}'/S^{1}\to \mc M_{g,1}.
\end{equation}

\begin{lem}\label {l1}
We have natural isomorphisms:
\begin{equation}\label{mg0'}
H_{S^{1}}^{*}(\widehat{\mc M}_{g})=H_{S^{1}}^{*}(\mc
M_{g,1}')=H^{*}(\mc M_{g,1}).
\end{equation}
The generator of the algebra $H_{S^1}(pt)=\mathbb{C}[u]$ acts on
$H^{*}(\mc M_{g,1})$ as multiplication by $-\psi$ where $\psi$ is
the first Chern class of the complex line bundle $K_{\pi}$ over
${\mc M}_{g,1}$ that over the point $(C,p)\in \mc M_{g,1}$ has a
fiber defined as the cotangent space to $C$ at the point $p$. (This
bundle can be interpreted as relative dualizing sheaf of fibration
${\mc M}_{g,1}\to {\mc {M}}_g.$)

\begin{proof}
Since $S^{1}$ acts on both $\mc M_{g,1}'$ and $\widehat{\mc M}_{g}$
freely and $\mc M_{g,1}'$ is homotopy equivalent to $\widehat{\mc
M}_{g}$, we have the natural identifications
\begin{equation}\label{mg1'}
H_{S^{1}}^{*}(\mc M_{g,1}')=H^{*}(\mc M_{g,1}'/S^{1})\cong
H^{*}(\widehat{\mc M}_{g}/S^{1})=H_{S^{1}}^{*}(\widehat{\mc M}_{g}).
\end{equation}
The map $\bar{F}'$ defined in (\ref{ID'}) is a homeomorphism which
identifies the cohomology:
\begin{equation}\label{mg1}
H^{*}(\mc M_{g,1}'/S^{1})\cong H^{*}(\mc M_{g,1}).
\end{equation}
By (\ref{mg1'}) and (\ref{mg1}), we proved (\ref{mg0'}).

To find the action of the generator of the algebra $H_{S^1}(pt)$ we
apply the general statement that in the case of free action of $S^1$
on $X$ the action of the generator on $H^*_{S^1}(X)=H^*(X/S^1)$ can
be described as multiplication on the first Chern class of the
circle bundle $X\to X/S^1$.
\end{proof}
\end{lem}

The moduli space $\widehat{\mc F}_{g,h}$ of quintuples
$(C,p,z,L,\phi)$ is homotopy equivalent to the moduli space $\mc
F_{g,h}'$ of quaduples $(C,p,v,L)$, where $(C,p,L)$ specifies an
element of $\mc F_{g,h}$ and $v$ is a tangent vector to $C$ at $p$.
Similarly, we have the following result:

\begin{lem} \label {l2}
There exists a natural isomorphism:
\begin{equation*}
H_{S^{1}}^{*}(\widehat{\mc F}_{g,h})\cong H^{*}(\mc F_{g,h}).
\end{equation*}
The generator of the algebra $H_{S^1}(pt)=\mathbb{C}[u]$ acts on
$H^{*}(\mc F_{g,h})$ as multiplication by $-\omega$ where $\omega$
denotes the first Chern class of line bundle over
$\mathcal{F}_{g,h}$ having the cotangent space $T_p^{*}$ to the
curve $C$ at $p$ as a fiber over $(C,p,L)\in \mathcal{F}_{g,h}$.
(This bundle can be regarded as the relative dualizing sheaf of the
forgetful map $\pi':\mc F_{g,h}\to\mc P_{g,h}$.)
\end{lem}

Notice that  the group $S^{1}$ acts on $\hs =L^{2}(S^{1})$  as the
group of rotations of  $S^1$; this action induces   an action
on $\Gr(\hs)$. It is easy to check that the Krichever map commutes
with the $S^1$-action on moduli spaces and on Grassmannian, hence it
induces a homomorphism on equivariant cohomology. Our goal is to
study this homomorphism.

\subsection{Topology of Sato Grassmannian}
Let us remind some basic facts about topology of finite-dimensional
Grassmannian $\Gr _{n,l}$ (of the space of $l$-dimensional complex
vector subspaces of $\mathbb{C}^n$.) The torus $T=(S^1)^n$ (as well
as the algebraic torus $\mb {T}^{al}=(\mathbb{C}^*)^n$) acts in natural
way on $\mathbb{C}^n$ and therefore on $\Gr _{n,l}$. (The torus acts
on $\mathbb{C}^n$ by means of linear transformations having vectors
of the standard basis $\{e_1,\dots,e_n\}$ as eigenvectors.) Fixed
points of the torus action on $\Gr _{n,l}$ are vector subspaces
$H_S$ spanned by subsets $S$ of the the set $\{e_1,\dots,e_n\}$
consisting of $l$ vectors. There exists a cell decomposition into
even-dimensional cells invariant with respect to torus action
(Schubert cells); these cells are in one-to-one correspondence with
fixed points. It follows that the Grassmannian is equivariantly
formal. This means that equivariant cohomology is a free module over
the cohomology of one-point set; cocycles dual to Schubert cells
(Schubert cocycles) constitute a basis of this module.  The
two-dimensional orbits of the torus action (orbits of  $\mb
{T}^{al}$ having complex dimension $1$) can be described in the
following way. Let us consider two fixed points of torus action
corresponding to subsets $S_1, S_2$ having $l-1$ common vectors.
Denote the  subspace spanned by vector $e_i+\lambda e_j$ and vectors
from $S_1\bigcap S_2$ by $V_{\lambda}$ (here $\lambda\in\mathbb{C}$,
$e_i\in S_1\setminus S_1\bigcap S_2,e_j\in S_2\setminus S_1\bigcap
S_2$). These subspaces form a  two-dimensional orbit of the torus
action. Applying GKM theorem one obtains the equivariant cohomology
ring of Grassmannian as a subring of the ring of functions on the
set of fixed points taking values in the polynomial ring $\mb C
[u_1,...,u_n]$.

The situation with Sato Grassmannian is similar. An
infinite-dimensional torus $\mathbb{T}$ acts on $\hs$ {\footnote {If
$(\dots,t_n,\dots)\in \mb T,$  $n\in \mathbb{Z}$, the corresponding
map of $\hs$ transforms a point $\sum a_ne_n$ into the point $\sum
t_na_ne_n$. Here $e_n=z^n$.The topology of $\mathbb{T}$ is specified
by the operator norm. The  cohomology ring  $H_{\mathbb{T} }^*(pt)$
can be considered as a subring of the ring of functions of infinite
number of variables $u_n$ where $n\in \mathbb{Z}$ (this follows from
the fact that every homomorphism of $S^1$ into $\mb T$ induces a map
$H_{\mathbb{T} }^*(pt)\to H_{S^1}^*(pt)$.)  One can give a precise
description of this subring, but we do not need this description.}}
and therefore on $\Gr(\hs)$ . The fixed points of this action are
subspaces $\hs _S$ spanned by vectors $z^j$ where $j\in S$. Such a
subspace belongs to $\Gr (\hs)$ iff $S\in\mathcal{S}$ where
$\mathcal{S}$ consists of subsets of $\mb Z$ that differ from $\mb
Z_{-}$ only by finite number of points, i.e. the symmetric
difference  $S\Delta\mb Z_{-}$ is a finite set . Two-dimensional
orbits of torus action correspond to pairs of subsets $S_1, S_2$
such that  one can go from one subset to another deleting and adding
one vector. The construction of such an orbit is similar to the
construction in finite-dimensional case. We can describe equivariant
cohomology classes of Grassmannian $\Gr_d(\hs)$ in terms of their
restriction to fixed points (to the points of the form $\hs _S$). We
prove that the GKM theorem can be applied to Sato Grassmannian. This
allows us to describe the ring $H_{\mb T}(\Gr_d(\hs))$ as a subring
of functions $\phi$ on $\mathcal{S}$ taking values in  the ring
$H_{\mathbb{T} }^*(pt)$ (we consider only functions with finite
support). Namely, if $S_1=(S_1\cap S_2 )\cup \{e_i\}, S_2=(S_1\cap
S_2)\cup \{e_j\}$ the difference $\phi (S_1)-\phi (S_2)$ should be
divisible by $u_i-u_j$. We are mostly interested in cohomology
$H_{S^1}(\Gr_d(\hs))$; it can be described as a ring of $\mb
{C}[u]$-valued functions $\phi$ on $\mathcal{S}$ such that the
difference $\phi (S_1)-\phi (S_2)$ is divisible by $u$. (We embed $S^1$  into $\mb T$ by the formula
$t_i=\lambda^{i}$ and therefore we should substitute $iu$ instead of
$u_i$. )

There exists a stratification of $\Gr (\hs)$ in terms of Schubert
cells: the Grassmannian can be represented as a disjoint union of
Schubert cells $\Sigma _S$; again these cells are in one-to-one
correspondence with fixed points (the fixed point $\hs _S$ belongs
to the cell $\Sigma _S$). Instead of a set $S$ one can consider a
decreasing sequence $(s_{i})_{i\geq 1}$ of elements of this set; it
is easy to check that for $n>>0$ we have $s_{n}=-n+d$ where $d$
stands for the index of $\hs_S.$ The complex codimension of Schubert
cell $\Sigma _S$ is given by the formula
\begin{equation*}
l(S)=\sum_{i=1}^{\infty}(s_{i}+i-d).
\end{equation*}
To construct a stratification of Grassmannian $\Gr(\hs)$ we notice
that every subspace $V\subset \hs$ that specifies a point of
Grassmannian has a canonical basis of the form
$e_n=z^{s_n}+\sum_{l\geq s_{n}+1}k_{nl}z^l$ where $(s_{i})_{i\geq
1}$ is a decreasing sequence and $k_{ns_{j}}=0$ for all $1\leq j<n$.
The Grassmannian is a union of sets labelled by sequences $(s_{i})$
that appear in the definition of canonical basis; these sets are
called Schubert cells, they will be denoted by $\Sigma_{S}$.
{\footnote{An equivalent definition of Schubert cell can be given in
the following way. For every $S\in \mathcal{S} $ we construct  a set
$U_S$ consisting of such elements $V\in \Gr (\hs)$ that the
projection $V\to \hs _S$ is an isomorphism. Then we can find a basis
of $V$ having the form $e_n=z^n+\sum_lk_{nl}z^l$ where $n \in S,
l\notin S$. To define $\Sigma _S\subset U_S$ we impose an additional
condition $k_{nl}=0$ for $l<n$. }} Notice that instead of sequences
$S=(s_{n})$ one can use partitions $\lambda=(\lambda_{n})$ where
$\lambda _n=s_n+n-d$ vanishes for $n>>0$. Given a sequence $S$ with
its corresponding partition $\lambda$, we also denote $\Sigma_{S}$
by $\Sigma_{\lambda,d}$ or simply $\Sigma_{\lambda}$ when the index
$d$ is specified.

The closure $\bar{\Sigma}_{S}$ of $\Sigma_{S}$ is called the
Schubert cycle with the characteristic sequence $S=(s_{i})$. It
defines a cohomology class in $H^{2l(S)}(\Gr(\hs))$ (Schubert
class). The Schubert cycle is $\mathbb{T}$-invariant, hence it
specifies an element of equivariant cohomology group
$H_{\mathbb{T}}^{2l(S)}(\Gr(\hs))$. This element (also called
Schubert class) will be denoted by the symbol $[\bar{\Sigma} _{S}].$
{\footnote {More precisely, if the sequence $S$ has index $d$ the
class $[\bar{\Sigma} _S]$ belongs to
$H_{\mathbb{T}}^{2l(S)}(\Gr_d(\hs)).$}} We will be mostly interested
in equivariant cohomology $H_{S^1}^{2l(S)}(\Gr(\hs))$, where $S^1$
stands for the subgroup of $\mathbb{T}$ corresponding to rotation
$z\to \lambda z$, but the statements of the next paragraph can be
generalized to any subtorus of $\mathbb{T}$.

One can prove that the Schubert cocycles specify a basis of cohomology of
Grassmannian. Similarly equivariant cohomology is a free module over
$H_{S^1}(pt)$ generated by equivariant Schubert classes.  The
multiplication of Schubert classes can be expressed in terms of
Schur functions , as in finite-dimensional case. More precisely, if
Schubert classes are labeled by partitions, the multiplication
formula for Schubert classes in the case of Sato Grassmannian is the
same as for finite-dimensional Grassmannian. The multiplication of
equivariant Schubert classes can be expressed in terms of shifted
Schur functions introduced in \cite {OO} ; see \cite {LS}.

It is easy to check that (equivariant) cohomology of $\Gr_d(\hs)$ is
a polynomial algebra generated by Schubert classes $c_r$
corresponding to sequences $S=(s_{j})$, where $s_{j}=1-j+d$ for
$1\leq j\leq r$ and $s_{j}=-j+d$ for $j\geq r+1$.\footnote{The
corresponding partitions  are $(1,\dots,1,0,\dots)=(1^{r})$ for
$r\geq 1$.} If we are working with equivariant cohomology we will
use notations $C_r$ for these Schubert classes.{\footnote {The
classes $c_r$ can be interpreted as Chern classes of
(infinite-dimensional) tautological vector bundle over $\Gr_d(\hs)$
(up to a factor $(-1)^r$). We do not use this interpretation,
because it does not work in equivariant case: equivariant Chern
classes are not well defined for tautological bundle.}}

The proof of these statements can be based on the results of \cite
{PS}. Following \cite {PS} we can consider the sequence $G_k\subset
\Gr (\hs)$ where the subspace $V\in \Gr(\hs)$ belongs to $G_k$ iff
$\hs _{-k}\subset V\subset \hs _{k}$ . (We use the notation $\hs _k$
for the subspace of $\hs$ spanned by $z^n$ with $n<k$.) It follows
from \cite {PS} that the homology and cohomology of $\Gr (\hs)$
coincide with homology (cohomology) of the union $\Gr_0$ of sets
$G_k$. It is easy to derive from this fact that the same is true for
equivariant  homology and cohomology. The space $\Gr_0$ admits a cell
decomposition consisting of invariant even-dimensional cells \cite
{PS}; this decomposition can be used to calculate (equivariant)
(co)homology and justify the above statements. A little bit
different proof is based on the remark that the homology of $\Gr_0$
can be represented as direct limit of homology groups of $G_k$; the
same is true for equivariant  homology. {\footnote{ More generally,
one can consider spaces $G_{kl}$ consisting of subspaces obeying
$\hs_l\subset V\subset \hs_k)$ and take the limit $k\to \infty, l\to
-\infty.$}} The remark that $G_k$ is homeomorphic to disjoint union
of finite-dimensional Grassmannians permits us to finish the proof.

\section {Cohomological properties of Krichever map}

The action of the Krichever map on the cohomology of Grassmanian
$\Gr_d(\hs)$ can be expressed in terms of lambda-classes. If we are
working with equivariant cohomology, we can get analogous results by
introducing the notion of equivariant lambda-classes.

Recall that the Hodge bundle $\mathbb{E}$ over moduli space
$\mathcal{M}_g$ is defined as a bundle having as a fiber over a
curve $C\in \mathcal{M}_g$ the space of all holomorphic
differentials on $C$. Replacing the space of holomorphic
differentials by the space of holomorphic $q$-differentials in the
definition of the Hodge bundle, we obtain a more general notion of
Hodge bundle $\mathbb{E}_q$. (For $q>1$ this is a bundle of
dimension $d_q=(2q-1)(g-1)$.) More rigorously we can define Hodge
bundle $\mathbb{E}_q$ as the pushforward of $q$-th power of relative
dualizing sheaf $K_{\pi}$ of forgetful map $\pi:\mathcal{M}_{g,1}\to
\mc {M}_g$.  (As we noticed this sheaf can be identified with  the complex line
bundle  over $\mathcal{M}_{g,1}$ that over the point $(C,p)\in \mc
M_{g,1}$ has a fiber defined as cotangent space to $C$ at the point
$p$.  The first Chern class of this bundle was denoted by $\psi$.)

Chern classes of Hodge bundle are called lambda-classes; they are
denoted by $\lambda _r$ (if we would like to emphasize that we are
working with $q$-differentials we use the notation $\lambda _r^q$).
For all other moduli spaces, we have a natural map onto
$\mathcal{M}_g$; taking pullback with respect to this map, we
construct Hodge bundles and lambda-classes on these spaces.

Hodge bundles over $\widehat{\mc M}_{g}$ are $S^1$-equivariant
bundles, hence we can define corresponding equivariant Chern
classes. They are called  equivariant lambda-classes and denoted by
$\Lambda _r$. The equivariant Chern classes of $\mb E_{q}$ are
denoted by $\Lambda _r^q$. Equivariant lambda classes are
equivariant cohomology classes of $\widehat{\mc M}_{g}$, or
equivalently, cohomology classes of ${\mc M}_{g,1}$ . If $\pi$ is
the natural projection ${\mc M}_{g,1}\to {\mc M}_g$ we can say that
$\Lambda _r^q=\pi ^*\lambda _r^q$.  To prove this fact we notice
that $\mathcal{M}_g$ can be considered as $S^1$-space and Hodge
bundle as an equivariant bundle over it if we assume that $S^1$ acts
 trivially. The equivariant Hodge bundle over $\widehat{\mc
M}_{g}$ can be regarded as pullback of Hodge bundle over
$\mathcal{M}_g$ with respect to natural projection $\widehat{\mc
M}_{g}\to\mathcal{M}_g$; this allows us to obtain $\Lambda _r^q$
from $\lambda _r^q$ considered as equivariant Chern class of a
bundle with trivial action of $S^1$.

The following theorem describes the behavior of equivariant
cohomology  classes $C_r$ with respect to the Krichever map.
(Notice, that $k_q^*(u)=-\psi$ as follows from Lemma \ref {l1}.)
It is sufficient to calculate Krichever map on the equivariant
cohomology ring of $\Gr _d(\hs)$ (classes $C_r$ generate this ring),
however, it is possible to calculate directly the image of Schubert
classes $[\bar{\Sigma} _{S}]$ (The result can be
expressed in terms of shifted Schur functions  defined in \cite
{OO}. This calculation will be published separately.)

\begin{thm}\label {Kr}
In the case $q>1$,
\begin{equation}\label{q1}
k_{q}^{*}C_{r}=(-1)^{r}\sum_{j+m=r}(-1)^{m}h_{m}(q,q+1,\dots,q+d_{q}-r)\psi^{m}\Lambda_{j}^{q}
\end{equation}
for all $1\leq r\leq d_{q}$ and
\begin{equation}\label{q1_1}
k_{q}^{*}C_{r}=(-1)^{r}\sum_{m+j=r}e_{m}(q-1,q-2,\cdots,q-r+d_{q}+1)\psi^{m}\Lambda_{j}^{q},
\end{equation}
if $r>d_{q}$.

Here $h_{m}(x_{1},x_{2},\dots)$  is the $m$-th complete symmetric
function in variables $\{x_{1},x_{2},\dots\}$ and
$e_{m}(x_{1},x_{2},\cdots)$ denotes the $m$-th elementary symmetric
function in variables $\{x_{1},x_{2},\cdots\}$.
\end{thm}
\begin{proof}
In order to calculate the homomorphism $k_q^*$  we introduce the
space $\Gr_{d}^{l}$ as the submanifold of $\Gr_d(\hs)$  consisting
of all $W$ such that the orthogonal projection $\pi_{l}:W\to
z^{-l}\hs_{-}$ is surjective.  It follows from this requirement that
there is an equivariant $(d+l)$-dimensional vector bundle $\mc
E_{l}$ over $\Gr_{d}^{l}$ whose fiber over $W$ is the kernel of the
projection $\pi_{l}:W\to z^{-l}\hs_{-}$.

The intersections of Schubert cells with  $\Gr_{d}^{l}$ form a
stratification of $\Gr_{d}^{l}$; the strata are also called Schubert
cells. For every Schubert cell in $\Gr_{d}(\hs)$ and sufficiently
large $l$ this cell is in general position with respect to
$\Gr_{d}^l (\hs)$; in other words the corresponding cell in
$\Gr_{d}^l (\hs)$ has the same codimension. (Recall that the
codimension of the Schubert  cell $\Sigma_{S}$ is determined by the
length $l(S)$). Denote the intersection of $\bar{\Sigma}_{(1^{r})}$
and $\Gr_{d}^{l}$ by $\bar{\Sigma}_{(1^{r}),l}$. The equivariant
cohomology class corresponding to  $\bar{\Sigma}_{(1^{r}),l}$ is
denoted by $C_{r,l}$. Since $\bar{\Sigma}_{(1^{r}),l}$ is in general
position with respect to $\Gr_{d}^{l}$ for $l>>0$ applying the
formula (\ref{int}) we obtain
\begin{equation}
\label{rl}
f_{l}^{*}C_{r}=C_{r,l}
\end{equation}
where $f_{l}^{*}:H_{T}^{*}(\Gr_{d})\to H_{T}^{*}(\Gr_{d}^{l})$ is
the  homomorphism  induced by the inclusion map
$f_{l}:\Gr_{d}^{l}\to \Gr_{d}$.

Denote $d=d_{q}$. (Here $d_q=(2q-1)(g-1)$ stands for the dimension
of Hodge bundle $\mathbb{E}_q$, $q>1.$). The Krichever locus
$k_{q}(\widehat{\mc M}_{g})$ lies in $\Gr_{d}^{l}$ for all $l\geq
0$. We obtain modified Krichever maps $^{l}k_{q}:\widehat{\mc
M}_{g}\to \Gr_{d}^{l}$ for all $l\geq 0$. Then $k_{q}=f_{l}\circ\
^{l}k_{q}$, where $f_{l}:\Gr_{d}^{l}\to \Gr_{d}$ is the inclusion
map. Hence we can compute $k_{q}^{*}:H_{T}^{*}(\Gr_{d})\to
H_{T}^{*}(\widehat{\mc M}_{g})$ composing homomorphisms in the
sequence
\begin{equation*}
\begin{CD}
H_{S^{1}}^{*}(\Gr_{d})@>f_{l}^{*}>>H_{S^{1}}^{*}(\Gr_{d}^{l})@>^{l}k_{q}^{*}>>H_{S^{1}}^{*}(\widehat{\mc
M}_{g}).
\end{CD}
\end{equation*}
Due to (\ref {rl}) it is sufficient to calculate
$^{l}k_{q}^{*}{C}_{r,l}$. To do this we express $C_{r,l}$ in terms
of equivariant Chern classes. The expression we need can be obtained
from general Kempf-Laksov formula (see  \cite {KL} or \cite {Ful},
Lecture 8), but we can use also simpler Porteous formula.

Let us construct a vector bundle $\underline{\hs}_{i,j}$ over
$\Gr_{d}^{l}$ as a bundle with total space
$\hs_{i,j}\times\Gr_{d}^{l}$. Here $\hs_{i,j}$ is the subspace of
$\hs$ spanned by $\{z^{m}:i\leq m\leq j\}$. We  define the action
of $S^{1}$ on this bundle by
\begin{equation}\label{qact}
(\lambda,f, W)\mapsto (\lambda^{-q}f(\lambda^{-1}z),\lambda (W))
\end{equation}
where $\lambda\in S^{1}$, $f\in \hs_{i,j}$ and $W\in \Gr_{d}^{l}$. (We define $\lambda (W)$ as a space of functions $f(\lambda^{-1}z)$ where $f(z)\in W.$)
We also define an $S^{1}$-action on $\mc E_{l}$ by (\ref{qact}) on
the fiber of $\mc E_{l}$. Then the bundles $\underline{\hs}_{i,j}$
and $\mc E_{l}$ are non-trivial equivariant bundles. The total
equivariant Chern class $c^T$(the sum of all equivariant Chern
classes) of $\underline{\hs}_{i,j}$ is given by the formula
\begin{equation*}
c^T(\underline{\hs}_{i,j})=\prod_{m=i}^{j}(1-(q+m)u).
\end{equation*}

\begin{lem}
\begin{equation}\label{equi_r}
C_{r,l}=(-1)^{r}c_r^{T}(\mc E_{l}-\underline{\hs}_{-l,d-r}).
\end{equation}
\end{lem}

Note that the class $c_{r}^{T}(\mc E_{l}-\underline{\hs}_{-l,d-r})$
is well-defined because $\mc E_{l}$ and $\underline{\hs}_{-l,d-r}$
are equivariant complex vector bundles of finite rank over
$\Gr_{d}^{l}$.

To prove this lemma we consider an equivariant bundle map $\mc
E_{l}\to \underline{\hs}_{-l,d-r}$ defined by means of orthogonal
projection of fibers. The cycle $\bar{\Sigma}_{(1^{r}),l}$ can be
considered as degeneracy locus of this bundle map (this is the locus
where the rank of the map of fibers is $\leq l+d-r$). This allows us
to apply the Porteous formula \cite {GH} to calculate the dual
cohomology class.

Let us denote by $\mc E_l^0$ the restriction of the bundle $\mc E_l$
to $\Gr_{d}^{0}$. The restriction of $\underline{\hs}_{-l,-1}$ to
$\Gr_d^{0}$ will be denoted by $\underline{\hs}_{-l,-1}^0$. There
exists an exact sequence of equivariant bundles
\begin{equation*}
0\to \mc E_0\to \mc E_l^0\to  \underline{\hs}_{-l,-1}^0\to 0.
\end{equation*}
This means that
\begin{equation*}
c^T(\mc E_l^0)=c^T(\mc E_0)c^T(\underline{\hs}_{-l,-1}^0).
\end{equation*}
Let us consider the case $1\leq r\leq d$.
Using the relation
$\underline{\hs}_{-l,d-r}=\underline{\hs}_{-l,-1}\oplus
\underline{\hs}_{0,d-r}$
we obtain that
\begin{equation*}
c^T(\underline{\hs}_{-l,d-r})=c^T(\underline{\hs}_{-l,-1})c^T
(\underline{\hs}_{0,d-r}).
\end{equation*}
We have
\begin{align*}
(-1)^{r}k_{q}^{*}C_{r}&=(-1)^{r}\ ^{l}k_{q}^{*}C_{r,l}=\
^{l}k_{q}^{*}c_r^{T}(\mc E_{l}-\underline{\hs}_{-l,d-r})=\
^{0}k_{q}^{*}\iota _l^*c_r^{T}(\mc E_{l}-\underline{\hs}_{-l,d-r})\\
&=\ ^{0}k_{q}^{*}c_r^{T}(\mc E_{l}^0-\underline{\hs}_{-l,d-r}^0)=\
^{0}k_{q}^{*}c_r^{T}(\mc E_0-\underline{\hs}_{0,d-r}^0).
\end{align*}
Here $\iota_l$ stands for the embedding $\iota _l:\Gr_{d}^{0}\to\Gr
_d^l$, $\ ^{l}k_{q}=\iota_l\ ^{0}k_{q}.$ It remains to notice that
$\mathbb{E}_q=\ ^{0}k_{q}^{*}\mc E_0$ (the bundle $\mathbb{E}_q$ is
the pullback of $\mc E_0$ as an $S^{1}$-equivariant bundle with
respect to the action (\ref{qact})). Hence $\Lambda _r^q=\
^{0}k_{q}^{*}c_r^T(\mc E_0).$ We obtain
\begin{equation*}
(-1)^rk_{q}^{*} C_{r}=\left (\frac{c^{T}(\mb
E_{q})}{\prod_{j=0}^{d-r}(1+(q+j)\psi)}\right)_{[r]}.
\end{equation*}
where $(x)_{[r]}$ stand for $2r$-dimensional component of cohomology class $x.$

Using the Cauchy's identity,
\begin{equation}\label{Cauchy}
\prod_{j=0}^{d-r}(1+(q+j)x)^{-1}=\sum_{m=0}^{\infty}(-1)^{m}h_{m}(q,q+1,\dots,q+d-r)x^{m},
\end{equation}
we obtain
\begin{equation*}
^{l}k_{q}^{*}c^{T}(\mc E_{l}-\underline{\hs}_{-l,d-r}) =
\sum_{j=0}^{d_{q}}\sum_{m=0}^{\infty}(-1)^{m}h_{m}(q,q+1,\dots,q+d_{q}-r)\psi^{m}\Lambda_{j}^{q}
\end{equation*}
which implies (\ref{q1}). In the case $r>d$, very similar arguments
lead to the relation:
\begin{equation*}
(-1)^rk_{q}^{*}C_{r}=\ ^{0}k_{q}^{*}c_r^{T}(\mc
E_0\oplus\underline{\hs}_{-(r-d)+1,-1}^0).
\end{equation*}
This relation implies (\ref{q1_1}).
\end{proof}

Applying the forgetful map $H_{S^{1}}^{*}(\Gr_{d}(\hs))\to
H^{*}(\Gr_{d}(\hs))$, we obtain:
\begin{cor}
\begin{equation*}
k_{q}^{*}c_{r}=(-1)^{r}\lambda_{r}^{q}
\end{equation*}
if $r\leq d_{q}$ and
\begin{equation*}
k_{q}^{*}c_{r}=0,
\end{equation*}
if $r>d_{q}$.
\end{cor}

Of course, it is easy to give an independent proof of these formulas
(for example, interpreting $c_r$ as Chern classes of
infinite-dimensional tautological vector bundle).

Mumford \cite {Mum} has shown how to relate lambda classes $\lambda
_r$ to kappa classes $\kappa _r=\pi_*\psi^{r+1}.$ The same method,
based on Grothendieck-Riemann-Roch theorem, can be used to calculate
$\lambda_{r}^{q}$ in terms of kappa classes.

\begin{thm}
The $r$-th  component of the Chern character of $\mb E_{q}$ is given by
\begin{equation*}
\ch_{r}\mb E_{q}=\frac{B_{r+1}(q)}{(r+1)!}\kappa_{r},
\end{equation*}
where $B_{n}(q)$ is the $n$-th Bernoulli polynomial in $q$. (The
Bernoulli polynomials $\{B_{n}(x)\}$  are defined by the generating
function $te^{xt}/(e^{t}-1)=\sum_{n=0}^{\infty}B_{n}(x)t^{n}/n!$).
\end{thm}

This formula was given in \cite {Bini}.

The expression of Chern
classes in terms of Chern character is well known (see for example
\cite {IG}).

The behavior of the equivariant cohomology with respect to the
Krichever map $k_{1}$ is described as follows:

\begin{thm}
\begin{equation}\label{q2}
k_{1}^{*}C_{r}=(-1)^{r}\sum_{j+m=r}(-1)^{m}h_{m}(1,2,\dots,g-r)\psi^{m}\Lambda_{j}
\end{equation}
for $r\leq g-1$ and
\begin{equation}\label{q3}
k_{1}^{*}C_{g}=(-1)^{g}\Lambda_{g}
\end{equation}
and
\begin{equation*}
k_{1}^{*}C_{g+1}=0
\end{equation*}
and if $r\geq g+2$, we have
\begin{equation}\label{q4}
k_{1}^{*}C_{r}=(-1)^{r}\sum_{m+j=r}(-1)^{m}e_{m}(1,2,\cdots,r-g-1)\psi^{m}\Lambda_{j}.
\end{equation}
\end{thm}
\begin{proof}
The Krichever locus $k_{1}(\widehat{\mc M}_{g})$ lies in
$\Gr_{d}^{l}$ for all $l\geq 1$. Consider the modified Krichever
maps $^{l}k_{1}:\widehat{\mc M}_{g}\to \Gr_{d}^{l}$ for $l\geq 2$.
We compute $k_{1}^{*}$ via
\begin{equation*}
\begin{CD}
H_{S^{1}}^{*}(\Gr_{d})@>f_{l}^{*}>>H_{S^{1}}^{*}(\Gr_{d}^{1})@>
^{l}k_{1}^{*}>>H_{S^{1}}^{*}(\widehat{\mc M}_{g}).
\end{CD}
\end{equation*}
Denote $\mc E_{l}^{1}$ the restriction of the bundle $\mc E$ to
$\Gr_{d}^{1}$. The restriction of $\underline{\hs}_{i,j}$ to
$\Gr_{d}^{1}$ is denoted by $\underline{\hs}_{i,j}^{1}$ for each
$i,j$. Then there exists an exact sequence of equivariant vector
bundles:
\begin{equation}\label{exact1}
0\to\mc E_{1}\to \mc E_{l}^{1}\to\underline{\hs}_{-l,-2}^{1}\to 0
\end{equation}
which gives
\begin{equation*}
c^{T}(\mc E_{l}^{1})=c^{T}(\mc
E_{1})c^{T}(\underline{\hs}_{-l,-2}^{1}).
\end{equation*}
For $1\leq r\leq g$, using the relation
$\underline{\hs}_{-l,g-1-r}=\underline{\hs}_{-l,-2}\oplus
\underline{\hs}_{-1,g-1-r}$, we obtain
\begin{equation*}
c^{T}(\underline{\hs}_{-l,g-1-r})=c^{T}(\underline{\hs}_{-l,-2})c^{T}(\underline{\hs}_{-1,g-1-r}).
\end{equation*}
We have
\begin{align*}
(-1)^{r}k_{1}^{*}C_{r}&=(-1)^{r}\ ^{l}k_{1}^{*}C_{r,l}=\
^{l}k_{1}^{*}c_{r}^{T}(\mc E_{l}-\underline{\hs}_{-l,g-1-r})=\
^{1}k_{1}^{*}\
\iota_{1,l}^{*}c_{r}^{T}(\mc E_{l}-\underline{\hs}_{-l,g-1-r})\\
&=\ ^{1}k_{1}^{*}c_{r}^{T}(\mc
E_{l}^{1}-\underline{\hs}_{-l,g-1-r}^{1})=(-1)^{r}\
^{1}k_{1}^{*}c_{r}^{T}(\mc E_{1}-\underline{\hs}_{-1,g-1-r}).
\end{align*}
Here $\iota_{1,l}$ stands for the embedding
$\iota_{1,l}:\Gr_{d}^{1}\to \Gr_{d}^{l}$, $^{l}k_{1}=\iota_{1,l}\
^{1}k_{1}$. Notice that $\mb E=\ ^{1}k_{1}^{*}\mc E_{1}$ and hence
$\Lambda_{r}=\ ^{1}k_{1}^{*}c_{t}^{T}(\mc E_{1})$. We obtain
\begin{equation*}
(-1)^{r}k_{1}^{*}C_{r}=\left(\frac{c^{T}(\mb
E)}{\prod_{j=-1}^{g-1-r}(1+(j+1)\psi)}\right)_{[r]}
\end{equation*}
which implies (\ref{q2}) and (\ref{q3}) by the Cauchy's identity. By
the exact sequence (\ref{exact1}), we have
\begin{equation*}
(-1)^{g+1}k_{1}^{*}C_{g+1}=\ ^{1}k_{1}^{*}c_{g+1}^{T}(\mc
E_{l}^{1}-\underline{\hs}_{-l,-2}^{1})c=\
^{1}k_{1}^{*}c_{g+1}^{T}(\mc E_{1})=c_{g+1}^{T}(\mb E)=0.
\end{equation*}
In the case $r>g+1$, very similar arguments give us
\begin{equation*}
(-1)^{r}k_{1}^{*}C_{r}=\ ^{1}k_{1}^{*}c_{r}^{T}(\mc E_{1}\oplus
\underline{\hs}_{-(r-g),-2}^{1}).
\end{equation*}
This result gives us (\ref{q4}).
\end{proof}

Similarly, forgetting about the equivariant structure, we obtain
\begin{cor}
\begin{equation*}
k_{1}^{*}c_{r}=(-1)^{r}\lambda_{r}
\end{equation*}
for $r\leq g$,
\begin{equation*}
k_{1}^{*}c_{r}=0,
\end{equation*}
for $r>g$.
\end{cor}

The moduli space $\mathcal{M}_g$ can be embedded in the moduli space
$\bar {\mathcal{M}} _g$ (Deligne-Mumford compactification). Similar
embeddings exist for other moduli spaces we considered (we allow
curves with simple double points, but the marked point should be
non-singular). The Krichever map $k_1$ can be
 extended to the moduli space $\widehat{\bar{\mc
M}}_{g}$, but this extension is not continuous (however, the  extension is continuous on the subspace consisting of irreducible curves). More generally, the map $k_1$ can be extended to the moduli space of irreducible Cohen-Macaulay curves with a disk around a non-singular point; this extension is continuous in appropriate topology. (This follows from the results of \cite {GS} and from the remark that  the dualizing sheaf of Cohen-Macaulay curve is torsion-free.) Our methods can be applied to the analysis of cohomological properties of the extended Krichever map.
%To prove this theorem we notice that for $q>1$ the Krichever map
%sends  $\widehat{\mc M}_{g}$ into  $\Gr^0_{d_q}(\hs)$. The Hodge
%bundle $\mathbb{E}_q$ on $\widehat{\mc M}_{g}$ can be described as a
%pullback by the Krichever map of the bundle $ \mathcal{L}_{d_q}$
%over $\Gr^0_{d_q}(\hs)$. (This is true also in the case when both
%bundles are regarded as $S^1$-equivariant bundles.) We see that the
%Krichever map transforms Chern classes of the bundle
%$\mathcal{L}_{d_q}$ into lambda-classes (both in conventional and
%equivariant case). Combining this statement with Proposition
%{\ref{good}} we obtain the theorem. In the case $q=1$ .... The
%theorem \ref {Kr} permits us to give a complete description of the
%homomorphism induced by the Krichever map in the non-equivariant
%case.

Let us define a vector bundle $\mb P$    on the moduli space $\mathcal{P}_{g,h}$ (on the
moduli space of pairs $(C,L)$) as a bundle having a fiber
 over a point $(C,L)$  that can be
identified with the space of holomorphic sections of $L$.\footnote
{The restriction of $\mb{P}$ to moduli spaces of curves embedded
into ${\mc P}_{g,h}$ by means of $q$-differentials coincides with
Hodge bundle.} (To guarantee the existence of such a vector bundle
we impose the condition $h>2g-2$; then one of the terms in
Riemann-Roch theorem vanishes and $\mb P$ is a bundle of rank
$d=h-g+1$.) The Chern classes of $\mb{P}$ are denoted by the symbol
$p_r$; we will use the same notation for their images in cohomology
of other moduli spaces that can be mapped in $\mathbb{P}$ in natural
way.

The classes $p_r$ are analogous to  lambda-classes $\lambda _r^q$.
The methods that were applied to calculate lambda-classes can be
used to compute $p_r$. It is easy to check that the bundle $\mathbb{
P}$ is a pushforward of a bundle on $\mathcal{F}_{g,h}$. This
bundle, denoted by $\mc {L}$, has a fiber $L_p$ over the point
$(C,p,L)\in \mathcal{F}_{g,h}$. (Here $L_p$ stands for the fiber of
$L$ over the point $p$.) We denote its first Chern class by
$\gamma.$

Recall that we denoted by  $\omega$ be the first Chern class of
line bundle over $\mathcal{F}_{g,h}$ having the cotangent space
$T_p^*$ to the curve $C$ at $p$ as a fiber over $(C,p,L)\in
\mathcal{F}_{g,h}$ (see Lemma \ref {l2}).
%(This bundle can be regarded as the relative
%dualizing sheaf of the forgetful map $\pi':\mc F_{g,h}\to\mc
%P_{g,h}$.)
Following \cite {Kaw} we define the generalized Mumford-Morita classes $m_{i,j}$ by
\begin{equation*}
m_{i,j}=\pi_{*}(\gamma^{i}\omega^{j})\in H^{i+j-1}(\mc F_{g,h}).
\end{equation*}
The Chern classes of $\mb P$ can be expressed in terms of the
generalized Mumford-Morita classes $m_{i,j}$.

\begin{thm}
The $k$-th Chern character of $\mb P$ is given by
\begin{equation}\label{GGR}
\ch_{k}\mb
P=\frac{1}{(k+1)!}m_{k+1,0}-\frac{1}{2(k!)}m_{k,1}+\sum_{j=1}^{[k/2]}\frac{B_{2j}}{(2j)!(k-2j)!}m_{k+1-2j,2j},
\end{equation}
where $\{B_{n}\}$ are Bernoulli numbers
$\displaystyle\frac{x}{e^{x}-1}=\sum_{n=0}^{\infty}B_{n}\frac{x^{n}}{n!}$.
\end{thm}
\begin{proof}
The proof is based on the Grothendieck-Riemann-Roch theorem applied to the  forgetful map $\pi':\mc F_{g,h}\to\mc
P_{g,h}$ and the bundle $\mc L$ on $\mc F_{g,h}$; it is
similar to Mumford's calculation of lambda classes. The Todd class
of the relative tangent sheaf $\mc T_{\pi'}$ of $\pi'$ is
\begin{equation*}
\Td \mc
T_{\pi'}=\frac{-\omega}{1-e^{\omega}}=1-\frac{1}{2}\omega+\sum_{j=1}^{\infty}\frac{B_{2j}}{(2j)!}\omega^{2j}.
\end{equation*}
Since $\ch \mc L=e^{\gamma}$, we find
\begin{equation*}
\ch\mc L\cdot\Td \mc
T_{\pi'}=\sum_{k=0}^{\infty}\frac{\gamma^{k}}{k!}-\frac{1}{2}\sum_{k=0}^{\infty}\frac{\gamma^{k}\omega}{k!}+\sum_{k=0}^{\infty}\sum_{j=1}^{\infty}\frac{B_{2j}}{k!(2j)!}\gamma^{k}\omega^{2j}.
\end{equation*}
The Grothendieck-Riemann-Roch theorem states that $\ch\mb
P=\pi_{*}'(\ch\mc L\Td \mc T_{\pi'})$. This gives us
\begin{equation*}
\ch\mb
P=\sum_{k=0}^{\infty}\frac{m_{k,0}}{k!}-\frac{1}{2}\sum_{k=0}^{\infty}\frac{m_{k,1}}{k!}+\sum_{k=0}^{\infty}\sum_{j=1}^{\infty}\frac{B_{2j}}{(2j)!k!}m_{k,2j}
\end{equation*}
which implies (\ref{GGR}).
\end{proof}

The bundle $\mb P$  can be considered as
$S^1$-equivariant bundle with respect to trivial action of $S^1$.
Let us denote by $\mc P$ an equivariant bundle constructed as a
pullback of $\mb P$ with respect to the forgetful map $\widehat{\mc
F}_{g,h}\to \mc {P}_{g,h}$ . The equivariant Chern
classes $P_r$ transforms into $p_r$ under the identification
$H_{S^{1}}^{*}(\widehat{\mc F}_{g,h})=H^{*}(\mc F_{g,h})$. (Notice
that the non-equivariant Chern classes of $\mc P$ are the classes
$p_r$ in $H^*(\widehat{\mc F}_{g,h})$.)

Now let us consider the Krichever map
\begin{equation}\label{Kri}
k:\widehat{\mc F}_{g,h}\to \Gr_{d}(\hs),
\end{equation}
where $d=h-g+1$ and $h>2g-2$. The Krichever locus $k(\widehat{\mc
F}_{g,h})$ lies in $\Gr_{d}^{l}(\hs)$ for all $l\geq 0$.

To study (\ref{Kri}), we consider the action of $S^{1}$ on the
vector bundle $\underline{\hs}_{i,j}$ and $\mc E_{l}$ defined by
\begin{equation}\label{aact}
\lambda\cdot (f,W)\mapsto (f(\lambda^{-1}z),\lambda (W)).
\end{equation}
Then both $\underline{\hs}_{i,j}$ and $\mc E_{l}$ are nontrivial
equivariant vector bundles. The total equivariant Chern class of
$\underline{\hs}_{i,j}$ is
\begin{equation*}
c^{T}(\underline{\hs}_{i,j})=\prod_{m=i}^{j}(1-mu).
\end{equation*}
Moreover, the equivariant vector bundle $\mc P$ of rank $d$ over
$\widehat{\mc F}_{g,h}$ is the pullback of the equivariant
$d$-dimensional bundle $\mc E_{0}$ over $\Gr_{d}^0 (\hs)$ via $k$
with respect to (\ref{aact}). It follows from \ref{l2} that
$k^{*}u=-\omega$.

To calculate the homomorphism induced by (\ref{Kri}) on the
(equivariant) cohomology we repeat the arguments used in the proof
of (\ref {Kr}). We obtain

\begin{thm}
For the equivariant case, we have
\begin{equation*}
k^*C_r=(-1)^{r}\sum_{j+m=r}(-1)^{m}h_{m}(1,2,\dots,d-r)\omega^{m}P_{j},
\end{equation*}
if $r\leq d$,
\begin{equation*}
k^{*}C_{r}=(-1)^{r}\sum_{m+j=r}(-1)^{m}e_{m}(1,2,\cdots,r-d-1)\omega^{m}P_{j},
\end{equation*}
if $r>d$. For the nonequivariant case, we have
\begin{equation*}
k^{*}c_{r}=(-1)^{r}p_{r}
\end{equation*}
if $r\leq d$,
\begin{equation*}
k^{*}c_{r}=0
\end{equation*}
if $r>d$.
\end{thm}

{\bf Acknowledments} We are indebted to Yu. Manin, M. Movshev,  M. Mulase, A.Okounkov, B. Osserman, F. Plaza-Martin , A. Polishchuk and V. Vologodsky for very useful comments.

\bibliographystyle{amsplain}
%\bibliography{sampartb}

\begin{thebibliography}{99}\large

\bibitem{AK} Arbarello, E., de Concini, C., Kac, V.G., Procesi,
C.,: Moduli Spaces of Curves and Representation Theory. Comm. Math.
Phys. 117, no. 1, 1-36 (1988).

\bibitem {Bini} Bini, G.,: Generalized Hodge Classes on the Moduli
Space of Curves. Beitr\"{a}ge Algebra Geom. 44, no.2, 559-565
(2003).


\bibitem {Ful} Fulton, W.,: Equivariant Cohomology in Algebraic
Geometry. Lecture Notes by D. Anderson (2007).


\bibitem{GH} Griffiths, P., Harris, J.,: Principles of Algebraic
Geometry. Wiley-Interscience, New York (1978).


\bibitem {GKM} Goresky, M., Kottwitz, R., MacPherson, R.,: Equivariant
Cohomology, Koszul Duality, and the Localization Theorem. Invent.
Math. 131, no. 1, 25-83 (1998).


\bibitem {KL} Kempf, G., Laksov, D.,: The Determinantal Formula of
Schubert Calculus. Acta Math. 132, 153-162 (1974).


\bibitem {Kaw} Kawazumi, N,: A Generalization of the Morita-Mumford
Classes to Extended Mapping Class Groups for Surfaces. Invent. Math.
131, no. 1, 137-149 (1998).


\bibitem {LS} Liou Jia-Ming, Schwarz, A.: Equivariant cohomology of infinite-dimensional Grassmannian and shifted Schur functions,  arXiv:1201.2554, to be published in  Mathematical Research Letters




\bibitem {Mum} Mumford, D.,: Towards an Enumerative Geometry of the
Moduli Space of Curves. Arithmetic and Geometry, vol. II, 271-328,
Progr. Math., 36, Birkh\"{a}user Boston, Boston, MA (1983).


\bibitem{MM} Mulase, M.,: Algebraic theory of the KP equations. Perspective in
mathematical physics, 151-217 (1994).

\bibitem{IG} Macdonald, I.G.,: Symmetric functions and Hall
polynomials. 2nd edition Clarendon Press, Oxford (1995).


\bibitem {OO} Okounkov, A., Olshanski, G.,: Shifted Schur Functions,
St. Petersburg Math. J. 9, 239-300 (1998),Shifted Schur Functions
II. The Binomial Formula for Characters of Classical Groups and its
Applications. Kirillov's Seminar on Representation Theory, 245-271,
Amer. Math. Soc. Transl. Ser. 2, 181, Amer. Math. Soc., Providence,
RI (1998).




\bibitem{PS} Pressley, A., Segal, G.,: Loop groups. Oxford Mathematical
Monographs. Oxford Science Publications. The Clarendon Press, Oxford
University Press, New York (1986).




\bibitem{GS} Segal, G., Wilson, G.,: Loop groups and equations of KdV type. Inst. Hautes
Etudes Sci. Publ Math. no. 61, 5-65 (1985).

\bibitem{SZ} Sen, A., Zwiebach, B.,: Quantum background independence of closed string field
theory.
Nuclear Phys. B 423, no 2-3, 580-630 (1994).

\bibitem{AS} Schwarz, A.,: Grassmannian and String theory. Comm. Math. Phys. 199, no. 1, 1-24 (1998).



\end{thebibliography}

\end{document}